\documentclass[aps,prl,showpacs,preprint,amsmath,amssymb,superscriptaddress]{revtex4}

\usepackage[]{graphicx}
\usepackage{booktabs}
\usepackage{amsmath}

\begin{document}
\title{Modelling of partially coherent radiation
based on the coherent mode decomposition}

\author{Andrej Singer}
\altaffiliation[email: ]{andrej.singer@desy.de}
\affiliation{Deutsches Elektronen-Synchrotron DESY, Notkestr. 85, D-22607 Hamburg, Germany}
\author{Ivan A. Vartanyants}
\altaffiliation[email: ]{ivan.vartaniants@desy.de}
\affiliation{Deutsches Elektronen-Synchrotron DESY, Notkestr. 85, D-22607 Hamburg, Germany}
\affiliation{National Research Nuclear University, "MEPhI", 115409 Moscow, Russia}

\begin{abstract}
We present a method for the propagation of partially coherent radiation using coherent mode
decomposition and wavefront propagation. The radiation field is decomposed into a sum of independent
coherent modes. Each mode is then propagated separately using conventional wavefront propagation techniques.
The summation of these modes in the plane of observation gives the coherence properties of the propagated radiation.
As an example, we analyze
propagation of partially coherent radiation transmitted through a single circular aperture.
\end{abstract}
\maketitle


\section{Introduction}
\label{sec:intro}  

With the construction of the third generation synchrotron sources (see for example \cite{Petra3})
partially coherent radiation in the x-ray range
has become available. The advent of the free-electron lasers
\cite{A2007,E2010} has tremendously increased
the amount of coherent flux, that can be used by experimentalists.
Techniques like coherent x-ray diffractive imaging (CXDI) \cite{MCK1999,PWV2006,C2006} explicitly
utilize the high coherent flux and
promise new insights in structural biology, condensed matter physics, magnetism and other correlated systems
\cite{CN2010}.

Careful planing of the coherence-based experiments is required, which means that the knowledge of
the beam properties at the experimental station is in high demand. Although many computational
methods have been developed to calculate the
beam profile at the sample position, most of them do not provide the coherence properties of the beam in the plane of
observation.
In addition, the majority of these calculations are based on
the ray tracing, or the wave propagation
approach and are valid only in the limit of incoherent or fully coherent radiation, respectively. At the same time,
partially coherent radiation that is mostly available nowadays at the 3rd generation synchrotron sources
and free electron lasers is not covered by these methods.

Here, we present a general approach, which can be applied to partially coherent wavefields, and
which is capable of calculating both the intensity profile of the beam as well as the transverse coherence
properties of the radiation at any position in the
beamline. It is based on the results of statistical
optics, where the radiation field is described in terms of correlation
functions. In this work we represent the correlation function of the radiation field as a sum of coherent
modes \cite{MW1995}. Each mode is propagated separately through the beamline optics
using a wave propagation technique. At the sample position
the correlation function is given as a sum of the propagated modes. The intensity profile
and the transverse coherence properties of the radiation can be readily extracted from this correlation function.
A similar approach was used in the frame of geometrical optics \cite{ZCS2005} and for advanced
phase retrieval methods in coherent imaging \cite{WWQ2009}.

We applied this technique to calculate the radiation properties of a Gaussian Schell-model source. As an example, source
parameters of the free electron laser in Hamburg (FLASH) were considered
and propagation of the partially coherent radiation through a pinhole was analyzed.

\section{Methods}
\subsection{Definitions}
In the theory of partial coherence the mutual coherence function, $\Gamma(\mathbf r_1,\mathbf r_2;\tau)$,
plays the dominant role. It describes the correlation
between two complex scalar values of the electric field, $E(\mathbf r;t)$,
at different points $\mathbf r_1$, $\mathbf r_2$ and at different
times $t$ and $t+\tau$ \cite{MW1995,G2000}
\begin{equation}
  \Gamma(\mathbf r_1,\mathbf r_2;\tau) = \langle E(\mathbf r_1; t)^*E(\mathbf r_2;t+\tau)\rangle_T,
  \label{eq:MCF}
\end{equation}
where the brackets $\langle\cdots\rangle_T$ denote the time average. It is assumed that
the averaging is performed over times $T$ that are much longer than the fluctuation time of the field \cite{F1}.
When we consider the propagation of the correlation functions in space, it is convenient to introduce the cross spectral
density function, $W(\mathbf r_1,\mathbf r_2;\omega)$, as the Fourier
transform of the mutual coherence function \cite{MW1995}
\begin{equation}
  W(\mathbf r_1,\mathbf r_2;\omega) = \int\Gamma(\mathbf r_1,\mathbf r_2;\tau)e^{i\omega\tau}\mbox d\tau,
  \label{eq:MI}
\end{equation}
where $\omega$ is the angular frequency of the radiation.
By definition, when the two points $\mathbf r_1$ and $\mathbf r_2$ coincide, the cross spectral density
represents the spectral density \cite{F2}
of the radiation field,
  $I(\mathbf r;\omega) = W(\mathbf r,\mathbf r;\omega).$
The normalized cross spectral density is known as the spectral degree of coherence
\begin{equation}
  \mu_{12}(\omega) =  \frac{W(\mathbf r_1,\mathbf r_2;\omega)}{\sqrt{I(\mathbf r_1;\omega)I(\mathbf r_2;\omega)}}
 ,\qquad 0 \le |\mu_{12}|\le1.
  \label{eq:CCF}
\end{equation}
The modulus of the spectral degree of coherence is often measured in interference experiments as
the contrast of the interference fringes.
To characterize the transverse coherence properties of the wavefield by a single number the normalized degree of
coherence can be introduced \cite{VS2010}
\begin{equation}
  \zeta(\omega) = \frac{\int \left|W(\mathbf r_1,\mathbf r_2;\omega)\right|^2\mbox d\mathbf r_1\mbox d\mathbf r_2 }
  {\left(\int I(\mathbf r;\omega)\mbox d\mathbf r\right)^2 }.
  \label{eq:NDC}
\end{equation}
\subsection{Mode decomposition and propagation of the correlation function}
It can be shown that under very general conditions the cross spectral density 
can be decomposed into a sum of statistically independent coherent modes \cite{MW1995}
\begin{equation}
  W(\mathbf r_1, \mathbf r_2;\omega) = \sum_n \beta_n(\omega) E_n^*(\mathbf r_1;\omega)E_n(\mathbf r_2;\omega),
  \label{eq:MD}
\end{equation}
where $\beta_n(\omega)$ are the eigenvalues and $E_n(\mathbf r;\omega)$ are the eigenfunctions of the integral equation
\begin{equation}
  \int W(\mathbf r_1,\mathbf r_2;\omega)E_n(\mathbf r_1;\omega) \mbox d\mathbf r_1= \beta_n(\omega)E_n(\mathbf r_2;\omega).
  \label{eq:FIE}
\end{equation}
In particular, the mode decomposition (\ref{eq:MD}) can be applied to planar secondary sources \cite{MW1995},
where the cross spectral density \cite{F3},
$ W(\mathbf r_1,\mathbf r_2)=W(\mathbf u_1,\mathbf u_2;z_0)$,
of the radiation field is given in the source plane with the transverse coordinates
$\mathbf u=(x,y)$. The coordinate $z$ is defined along the optical axis and the position of the source is at $z_0=0$.

The mode decomposition is convenient in the analysis of the propagation of partially coherent radiation, when
only a small number of modes is required to describe the cross spectral density.
The coherent modes can be propagated separately along the optical axis and the cross spectral density function,
$W(\mathbf u_1,\mathbf u_2;z)$,
can be calculated at any position $z$ along the optical
axis via eq. (\ref{eq:MD}) replacing the modes $E_n(\mathbf u,z_0)$ by
the propagated modes $E_n(\mathbf u,z)$.
The eigenvalues, $\beta_n$, remain unchanged during the  propagation.

The propagation of the field for each mode over a distance $z$ along the optical axis
can be performed by a wave propagation technique.
In the case of free space propagation this can be done utilizing the Huygens-Fresnel principle \cite{BW1999}
\begin{equation}
  E_n(\mathbf u,z)=\int P_z(\mathbf u,\mathbf u') E_n(\mathbf u',z_0)\mbox d\mathbf u',
  \label{eq:propagation}
\end{equation}
where $E_n(\mathbf u',z_0)$ is the wavefield of the mode in the source plane and $E_n(\mathbf u,z)$ is the propagated mode at
position $z$. The propagator $P_z(\mathbf u,\mathbf u')$ is given by
\begin{equation*}
  P_z(\mathbf u,\mathbf u')=\frac{k}{2\pi i}\frac{e^{ikr}}{r}\chi(\theta).
\end{equation*}
where $k=\omega/c$ is the wavevector, $r$ is the distance between the points $(\mathbf u',z_0)$ and $(\mathbf u,z)$, $\theta$ is the
angle between the line joining $(\mathbf u',z_0)$ to $(\mathbf u,z)$ and the optical axis, and
$\chi(\theta)$ is the obliquity factor with $\chi(0)=1$ and $0\le \chi(\theta)\le1$.
In the paraxial approximation, when small angles $\theta$ are considered, the propagator reduces to
\begin{equation*}
  P_z(\mathbf u-\mathbf u')=\frac{k}{2\pi iz}\exp\left[\frac{ik}{2z}(\mathbf u-\mathbf u')^2\right].
\end{equation*}

In general, the propagation of the partially coherent radiation through an arbitrary
arrangement of the optical elements in a beamline can be performed in the following steps
\begin{enumerate}
  \item Decomposition of  the cross spectral density, $W(\mathbf u_1,\mathbf u_2;z_0)$,
    of the source into coherent modes $E_n(\mathbf u,z_0)$ according to eq. (\ref{eq:MD}).
  \item Propagation of all modes $E_{n}(\mathbf u,z_0)$ from the source plane through the optical elements to the
    observation
    plane using a wave propagation technique.
    For example, equation (\ref{eq:propagation}) can be used to propagate all modes from the
    source to the first optical element at the position $z$. For thin optical elements
    the transmitted modes are given by $E^{out}(\mathbf u,z)=T(\mathbf u)E(\mathbf u,z)$,
    where $T(\mathbf u)$ is the transmission function. At the next step the transmitted modes
    are propagated to the next optical element using eq. (\ref{eq:propagation}).
  \item Finally, after performing the previous step for all optical elements present in the beamline
    each mode is calculated in the plane of observation,
    and the cross spectral density, $W(\mathbf u_1,\mathbf u_2;z)$, is determined by eq. (\ref{eq:MD}).
\end{enumerate}
\begin{figure}[t]
\centering\includegraphics[width=7cm]{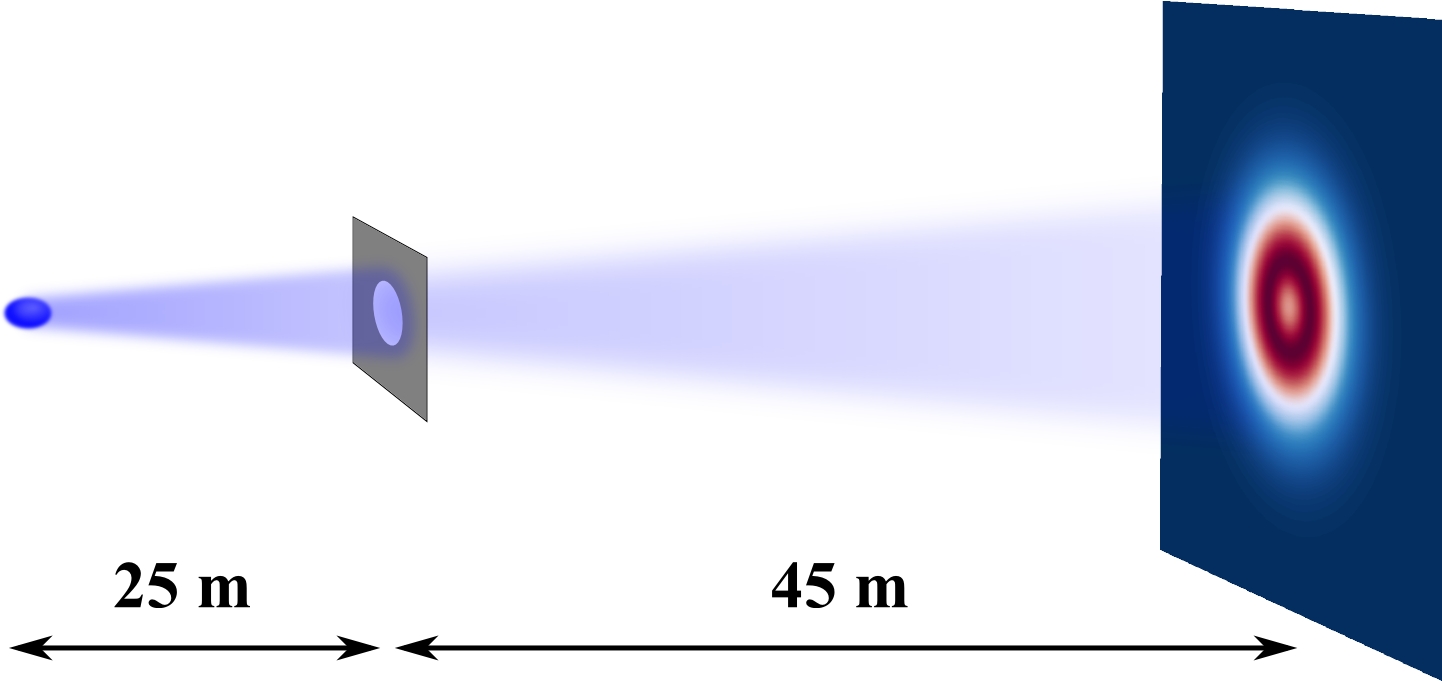}
\caption{Partially coherent radiation is generated at the source. The circular
aperture is positioned 25 m downstream from the source. The coherence properties of the radiation are analyzed
45 m downstream from the aperture.}
\label{fig:1}
\end{figure}
\begin{figure}[t]
\centering\includegraphics[height=10cm]{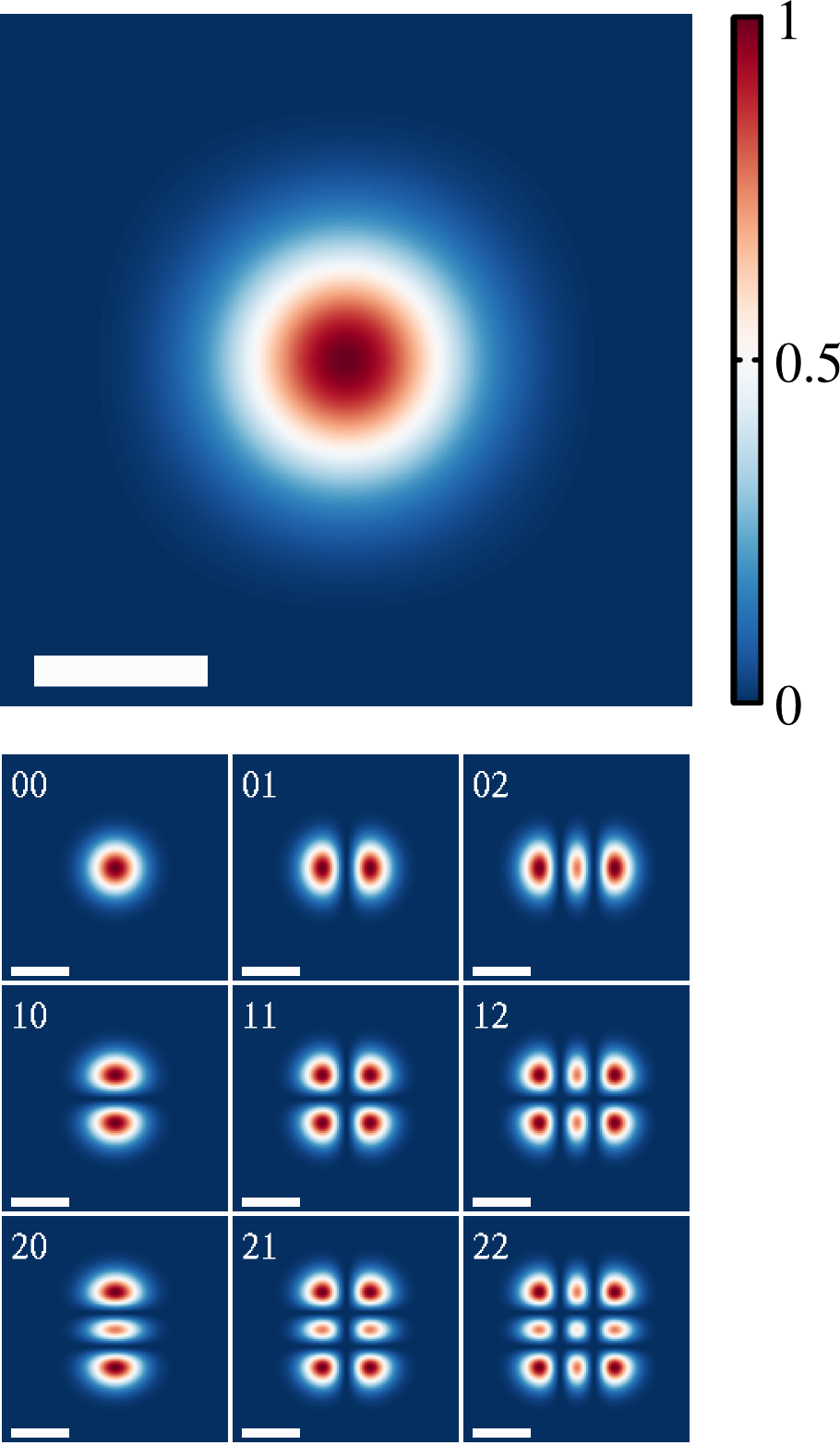}
\caption{(top) The total intensity distribution at the source position.
(bottom) The intensity distribution of the nine lowest modes
at the source position. The length of the white scale bar is 150 $\mu$m.}
\label{fig:2}
\end{figure}
\begin{figure}[tb]
\centering\includegraphics[height=10cm]{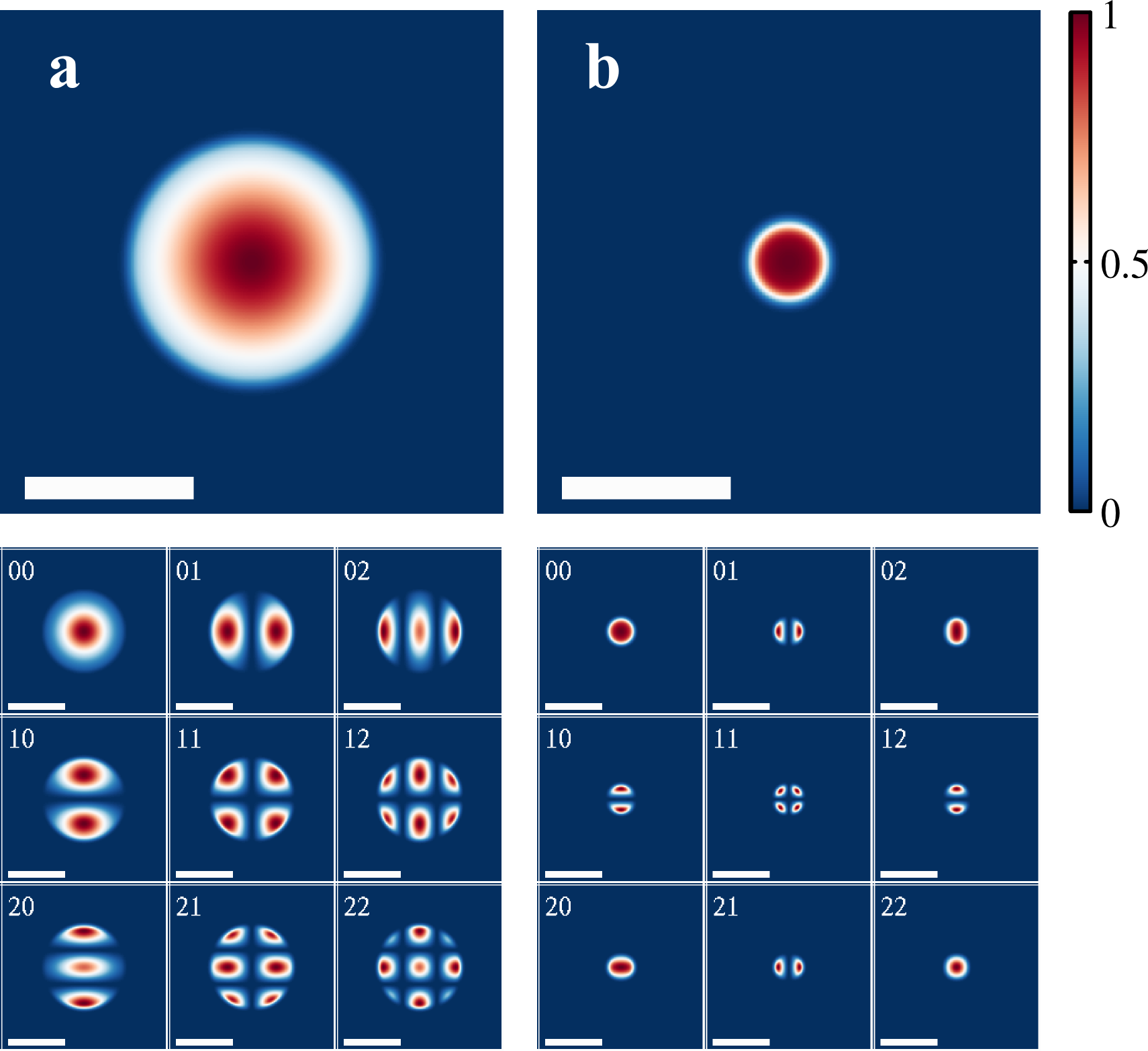}
\caption{(top) The total intensity distribution
behind a circular aperture with (a) 3mm and (b) 1 mm diameter. (bottom)
The intensity of the lowest nine modes behind the aperture.
The length of the scale bar is 2 mm.}
\label{fig:3}
\end{figure}
\begin{figure}[tb]
\centering\includegraphics[height=10cm]{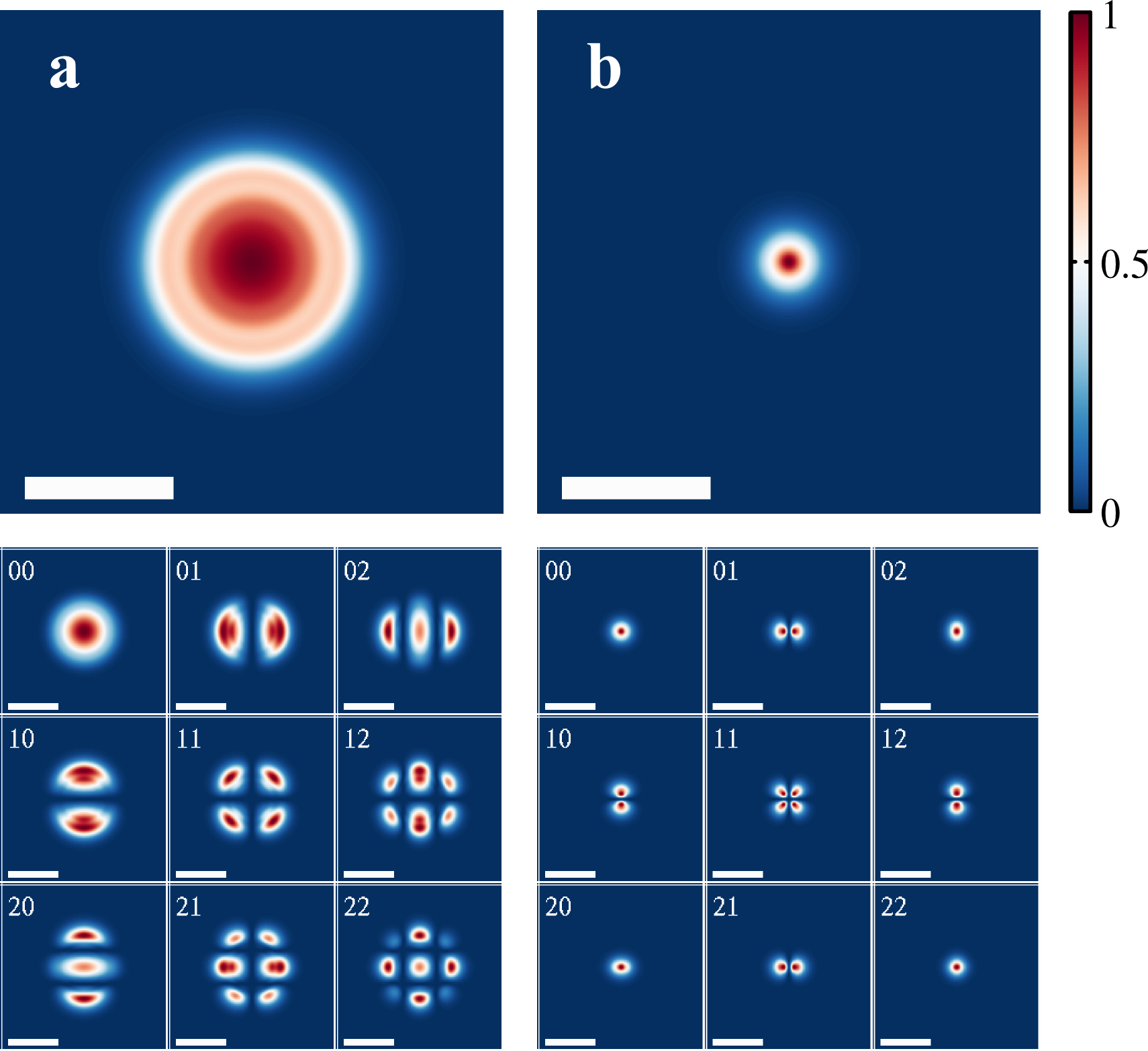}
\caption{(top) The total intensity distribution
in the plane of observation calculated for a circular aperture, (a) 3mm and (b) 1 mm in diameter. (bottom) The
intensity distribution of the lowest nine modes in the plane of observation.
The length of the scale bar is 5 mm.}
\label{fig:4}
\end{figure}

\subsection{Gaussian Schell-model}
A useful model to describe the radiation properties of partially coherent sources is the
Gaussian Schell-model \cite{MW1995}. This model has been
applied for the analysis of the radiation field generated by optical lasers \cite{G1980}, third
generation synchrotron sources \cite{VS2010} and x-ray free-electron lasers \cite{SVK2008,R2011,VSM2011}.
In this model the cross spectral density in the source plane, $W(\mathbf u_1,\mathbf u_2;z_0)$, is given by
\begin{equation}
  W(\mathbf u_1,\mathbf u_2;z_0)=\exp\left( -\frac{x_1^2+x_2^2}{2\sigma_x^2}-\frac{y_1^2+y_2^2}{2\sigma_y^2} \right)
\exp\left( -\frac{(x_2-x_1)^2}{2\xi_x^2}-\frac{(y_2-y_1)^2}{2\xi_y^2} \right),
\label{eq:GSM}
\end{equation}
where $\sigma_x,~\sigma_y$ is the source size and $\xi_x,~\xi_y$ is the transverse coherence length of the source in
the horizontal ($x$) and vertical ($y$) direction, respectively.
Due to the symmetry of the Gaussian Schell-model the total cross spectral density at the source
factorizes into the horizontal and vertical components
\begin{equation}
    W(\mathbf u_1,\mathbf u_2;z_0)=W(x_1,x_2;z_0)\cdot W(y_1,y_2;z_0).
    \label{eq:separation}
\end{equation}
The modes $E_n$  and their corresponding contributions $\beta_n$ can be found for each direction separately
$$W(x_1,x_2;z_0) = \sum_n \beta_n^{x}
E_n^*(x_1;z_0)E_n(x_2;z_0),$$
and a similar expression is valid for the $y$ direction.
The analytical solution of the integral equation (\ref{eq:FIE}) for the Gaussian Schell-model
in each direction is known in the form of the Gaussian Hermite-modes. The eigenvalues in this model
are described by a power law \cite{MW1995}.
The total cross spectral density is given by
\begin{equation}
  W(\mathbf u_1,\mathbf u_2;z_0)=\sum_{n,m}\beta_{nm}
  E_{nm}^*(\mathbf u_1;z_0)E_{nm}(\mathbf u_2;z_0),
  \label{eq:MD1}
\end{equation}
where $E_{nm}(\mathbf u;z_0)=E_n(x;z_0)\cdot E_m(y;z_0)$ are the well known laser resonator TEM$_{nm}$ modes \cite{G1983}
and $\beta_{nm} = \beta^x_n\beta^y_m$.
The total radiated intensity can be calculated by
\begin{equation}
  I(\mathbf u;z_0)=W(\mathbf u,\mathbf u;z_0)=\sum_{n,m}\beta_{nm}|E_{nm}(\mathbf u; z_0)|^2.
  \label{eq:int}
\end{equation}

We want to note here that our approach is not limited to the Gaussian Schell-model. If the cross spectral density of the
field, $W(\mathbf u_1,\mathbf u_2;z_0)$, at the source is known, the corresponding
modes $E_n$ and their contributions $\beta_n$ can be calculated from the integral equation (\ref{eq:FIE}) (see for
example Ref. \cite{FQT2009}).

\section{Propagation of partially coherent radiation through a circular aperture}
We simulated the propagation of the partially coherent radiation generated by a Gaussian Schell-model source
through a circular aperture (see Figure \ref{fig:1}). As source parameters we used values
reported  for  FLASH \cite{A2007}  operating at a wavelength of 13.5 nm:
the full width at half maximum (FWHM) source size is $160 \mu$m and the FWHM divergence of the beam is 90 $\mu$rad.
In the frame of the Gaussian Schell-model (see eq. (\ref{eq:GSM})) these values
correspond to $\sigma_x=\sigma_y=68~\mu$m and $\xi_x=\xi_y=62~\mu$m \cite{SVK2008}.
For these parameters the contribution of each mode at the source position
is given by the power law $\beta_{nm} = 0.41^{n+m}$. Modes with a contribution of less than 1 \%
were neglected in our simulations. In total, 21 modes with $n+m\le5$, where $n=0,1,\ldots 5,~m=0,1,\ldots5$,
were used in the calculations presented here.
The intensity distribution of the total field and the nine lowest modes  at the source position
are shown in Figure \ref{fig:2}.

We considered the following geometry in our simulations:
the pinhole is positioned 25 m downstream from the source and the radiation properties are analyzed
70 m downstream from the source (see Figure \ref{fig:1}). Such an arrangement is typical for experiments
performed in the unfocused beam at FLASH.
We applied the general scheme of propagation of partially coherent radiation described earlier for this
experimental geometry. The propagation in the free space was performed using equation (\ref{eq:propagation})
with the propagator in the paraxial approximation.
Different pinholes with the diameter from 5 mm to 1 mm were analyzed.
The pinhole transmission function, $T(\mathbf u)$, was defined as
    \[T(\mathbf u)=\begin{cases}
      1 & \textrm{for}\quad |\mathbf u|<d/2 \\
      0 & \textrm{elsewhere}
    \end{cases},
    \]
where $d$ is the diameter of the pinhole. It was convolved with a 200 $\mu$m wide (FWHM) Gaussian function
to account for imperfections of the pinhole edges.
The intensity distribution of the total
beam and the lowest modes behind 3 mm and 1 mm pinholes
are presented in Figure \ref{fig:3}. Figure \ref{fig:4} shows the same in the plane of observation.

One readily sees in Figure \ref{fig:3} (a) that in the case of the 3 mm pinhole the first modes, which dominate the radiation field, are
just slightly affected by the aperture. The 1 mm pinhole, however, substantially cuts
all modes, including the dominant ones (see Figure \ref{fig:3} (b)).
The intensity distribution of each mode in the plane of observation is similar to the intensity distribution leaving the
pinhole. Additional intensity modulations due to the scattering on the edges of the aperture are observed
in Figure \ref{fig:4} (a).
This can be attributed to the Fresnel diffraction effects, which are stronger for sharper pinhole boundaries.
The Fresnel number, $d^2/(\lambda L)$, where $\lambda$ is the
wavelength of the radiation and $L$ is the distance from the pinhole to the detector,
is 15 for the 3 mm pinhole and 1.6 for the 1 mm pinhole in this experimental geometry. We want to note
here that due to Fresnel diffraction, small variations in the propagation distances might change these intensity
modulations significantly.

\begin{figure}[tb]
\centering\includegraphics[height=12cm]{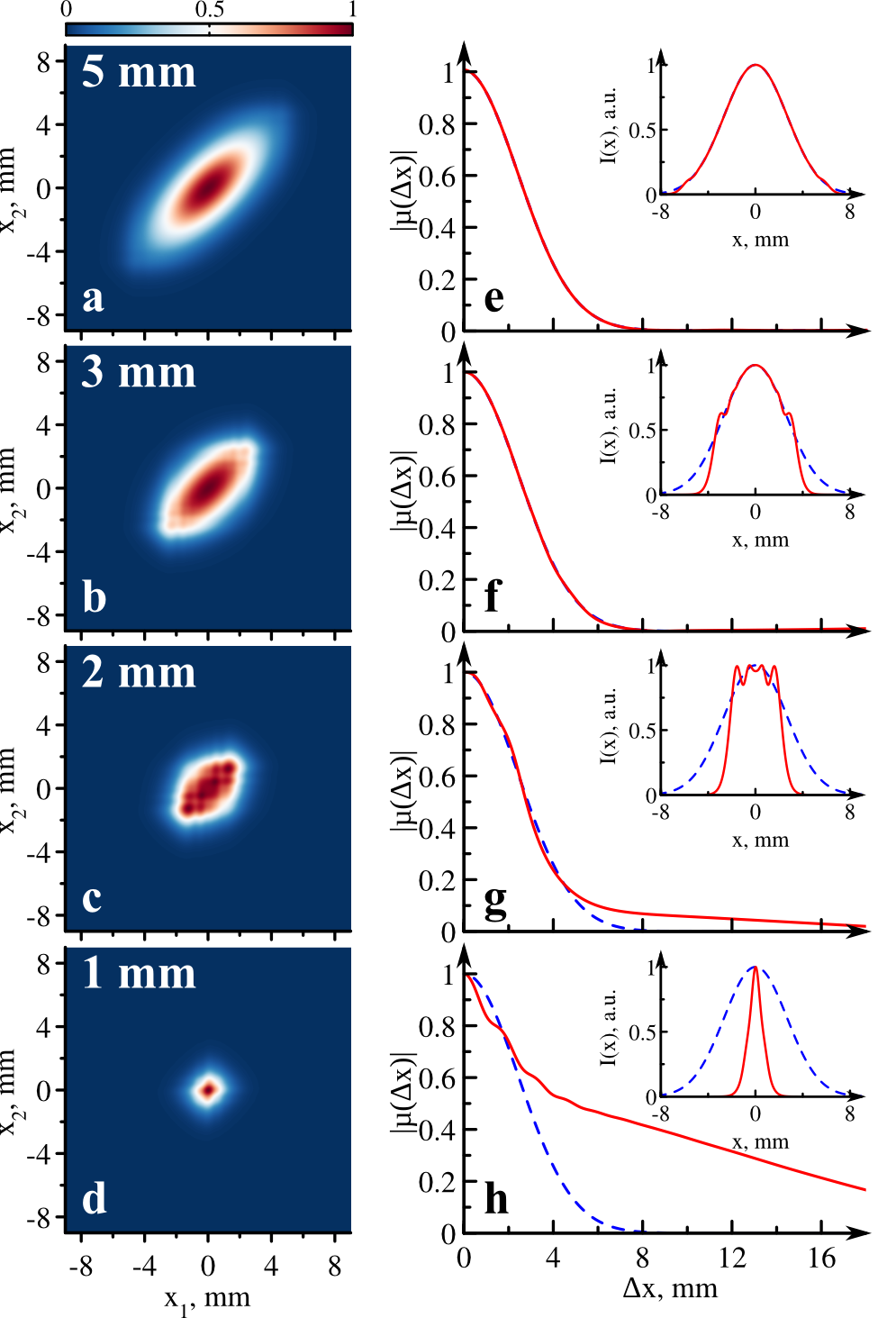}
\caption{Transverse coherence properties of the radiation in the observation plane. (a-d) The modulus of the cross spectral
density $|W(x_1,x_2;z)|$.
(e-h) The modulus of the spectral degree of coherence $|\mu(\Delta x)|$ as
a function of the separation $\Delta x$. Simulations were performed with the pinhole diameters of
(a,e) 5 mm, (b,f) 3 mm, (c,g) 2 mm, and (d,h) 1 mm.
The insets in (e-h) show the intensity distribution $I(x)$ in the horizontal direction.
Red solid lines show the simulations with the presence of the circular aperture. The blue dashed lines
show the same functions obtained without the pinhole.}
\label{fig:6}
\end{figure}

Finally, the cross spectral density, $W(\mathbf u_1,\mathbf u_2;z)$, was determined in the plane of observation.
We present here results for the horizontal direction
\begin{equation}
  W(x_1,x_2;z)=\sum_{nm}\beta_{nm}E_{nm}^*(x_1,y_1=0;z)E_{nm}(x_2,y_2=0;z).
\label{eq:csd}
\end{equation}
Due to the symmetry of our scattering geometry, the same result is obtained in the vertical direction.
The modulus of the cross spectral density, $|W(x_1,x_2;z)|$, as a function of the transverse positions,
$x_1$ and $x_2$, is shown in Figure \ref{fig:6} (a-d) for different apertures.
The modulus of the spectral degree of coherence $|\mu(\Delta x)|=|\mu(-\Delta x/2,\Delta x/2)|$
as a function of the separation $\Delta x$ for the same apertures is presented in Figure \ref{fig:6} (e-h) (red solid line).
This calculation corresponds to the measurements of the contrast in a double pinhole experiment
with varying pinhole separation $\Delta x$ and the center of the double pinhole positioned at the optical axis of the beam.
For comparison, the spectral degree of coherence, $|\mu(\Delta x)|$, for the same geometry and source
parameters, but without the aperture is shown by the blue dashed line. The intensity profiles, $I(x)$, for the
same aperture sizes are shown in the insets of Figure \ref{fig:6} (e-h).

\begin{table}[b]
\caption{The transmitted photon flux, $P$, and the
normalized degree of coherence, $\zeta$,  behind the aperture. Four different
pinhole diameters are analyzed. }
  \centering
  \begin{tabular}{c|cc}
    \toprule
   no pinhole & $P= 100\%$ &$\zeta = 18 \%$ \\
   \midrule
   5 mm       & $P= 97\%$  &$\zeta = 19 \%$ \\
   3 mm       & $P= 69\%$  &$\zeta = 29 \%$ \\
   2 mm       & $P= 39\%$  &$\zeta = 46 \%$ \\
   1 mm       & $P= 10\%$  &$\zeta = 78 \%$ \\
   \bottomrule
\end{tabular}
\label{table:1}
\end{table}
As a result of our simulations, we notice that the 5 mm pinhole does not affect the transmitted radiation.
The intensity profile $I(x)$ as well as the modulus of the spectral degree of coherence $|\mu(\Delta x)|$ in the observation plane
calculated with and without the pinhole are the same.
For the smaller pinhole diameters of 3 mm and 2 mm the size of the beam decreases, but the
modulus of the spectral degree of coherence is not significantly altered.
Our simulations suggest, that in the present geometry down to the pinhole sizes of 2 mm no significant changes
in the coherence properties of the beam in the observation plane are expected. Only for the smallest pinhole size of
1 mm the values of $|\mu(\Delta x)|$ are significantly enhanced at large separations (see Figure \ref{fig:6} (h)). In
this case the spectral degree of coherence cannot be described by a single Gaussian function (compare to results of
Ref. \cite{LPP2003}). However, we should note here that these separations are much larger than the beam size and will
be difficult to access in a real experiment.

We analyzed as well the normalized degree of coherence $\zeta$ (\ref{eq:NDC})
and the available photon flux $P$ behind each pinhole (see Table 1).
The transmitted photon flux was calculated by the equation
$P = \int I_{out}(\mathbf u)\mbox d\mathbf u/\int I_{in}(\mathbf u)\mbox d\mathbf u$, where
$I_{in}(\mathbf u)$ and $I_{out}(\mathbf u)$ are the intensity (see eq.
(\ref{eq:int})) incident on and behind the aperture.
If no pinhole is present in the beamline, then obviously the photon flux $P=100~\%$ and the normalized
degree of coherence has the
value determined by the source parameters $\zeta=18~\%$.
Results from the Table 1 show that the normalized degree of coherence, $\zeta$, is significantly increased
for the smaller pinholes. However,
this happens at a loss of the transmitted photon flux $P$.
It is interesting to note that the product $P\cdot\zeta$, which may be considered as the amount of the
coherent photon flux, is
about $20~\%$ after transmission through the larger pinholes.
It drops down to a value of only about $10~\%$ for the 1 mm pinhole. In the latter case
almost
a fully coherent beam , $\zeta=78~\%$, is achieved with 10 \% of the transmitted radiation.

\section{Conclusions}
In conclusion, we have presented a computational method, that allows the calculation of
the transverse coherence properties as well as the beam intensity profile of partially coherent radiation
at any position in the beamline. Our approach can be easily
implemented, since it is based on the wave front propagation, for which several powerful computational methods are
already developed.
The important extension to the conventional wave propagation methods is the consideration of all contributing modes.

We have applied this method to describe the propagation of partially coherent
radiation through a circular aperture. Parameters typical for FLASH were used in the simulations.
Our analysis shows, that the presence of the pinhole does not increase
the transverse coherence length substantially, even if the beam is significantly cut by the aperture.
At the same time, the normalized degree of coherence is significantly increased for smaller pinholes
at the expense of the available photon flux.
A straightforward  extension to our approach would be its application to a
beamline with a number of optical elements.
Our simulations suggest that a careful calculation is important to have a realistic picture
of the beam properties in the observation plane for a concrete realization
of the optics in the beamline.
An interesting question, for example, would be how imperfections
of the optical elements affect the transverse coherence length of the radiation.

This approach can be effectively used to describe the radiation at free electron lasers, which
is predominantly coherent and contains a small number of statistically independent modes.
It can be also applied to third generation synchrotron sources, which are highly coherent in the vertical
direction and less coherent in the horizontal direction \cite{VS2010}.
However, in this case a significantly larger number of modes has to be taken into account.

We would like to thank Edgar Weckert for his interest and support of this work and Ulf Lorenz
for his careful reading of the manuscript.
Part of this work was supported by BMBF Proposal 05K10CHG ''Coherent
Diffraction Imaging and Scattering of Ultrashort Coherent Pulses with Matter'' in the framework of the German-Russian
collaboration ''Development and Use of Accelerator-Based Photon Sources''.


\begin{thebibliography}{99}
\bibitem{F1} For pulsed sources
like free-electron lasers this assumption means that the pulse duration is much longer than the temporal coherence time.
\bibitem{F2} For the narrow-bandwidth light, that will be considered in the following,
it corresponds to the intensity distribution at the average frequency $\bar\omega$.
\bibitem{F3} In all equations below, we omit $\omega$ for brevity.
\end{thebibliography}
\end{document}